\documentclass[]{aa}
\usepackage{txfonts}
\usepackage{graphicx}
\usepackage{natbib}
\bibpunct{(}{)}{;}{a}{}{,}

\begin{document}

\title{Mapping ices in protostellar environments on 1000 AU scales}
\subtitle{Methanol-rich ice in the envelope of Serpens SMM 4
\thanks{Based on observations obtained at the European Southern Observatory, Paranal, Chile, within the observing program 71.C-0252(A).}}

\author{K. M. Pontoppidan\inst{1} \and E. F. van Dishoeck\inst{1} \and E. Dartois\inst{2}}
\institute{Leiden Observatory, P.O. Box 9513, NL-2300 RA Leiden, The Netherlands \and
	 Institut d'Astrophysique Spatiale, B{\^a}t. 121, Universit{\'e} Paris XI, 91405 Orsay Cedex, France}

\offprints{K. M. Pontoppidan \email{pontoppi@strw.leidenuniv.nl}}

\date{Received / Accepted}

\abstract{We present VLT-ISAAC $L$-band spectroscopy toward 10 stars in SVS 4, a $30\arcsec\times45\arcsec$ dense cluster of pre-main sequence stars deeply embedded in the Serpens star forming cloud. The ISAAC spectra are combined with archival imaging from UKIRT and ISOCAM to derive accurate extinctions toward the SVS 4 stars. The data are then used to construct a spatial map of the distribution of ice in front of the cluster stars with an average angular resolution of 6\arcsec or 1500\,AU, three orders of magnitude better than previous maps. We show that water ice is present throughout the region and confirm the presence of methanol ice with an abundance of up to 25\% relative to water. It is shown that methanol ice maintains a very high abundance relative to H$_2$ throughout SVS 4, but drops by at least an order of magnitude only $75\arcsec$ away from SVS 4. The maps indicate that some of the lines of sight toward the SVS 4 stars pass through the outer envelope of the class 0 protostar SMM 4. The abundance of water ice relative to the refractory dust component shows a sudden increase by 90\% to $(1.7\pm0.2)\times 10^{-4}$ relative to H$_2$ at a distance of 5\,000 AU to the center of SMM 4. The water ice abundance outside the jump remains constant at  $(9\pm1) \times 10^{-5}$. We suggest that this is an indication of a significantly enhanced ice formation efficiency in the envelopes of protostars. The depletion of volatile molecules in the envelope of SMM 4 is discussed. In particular, it is found that up to 2/3 of the depleted
CO is converted into CO$_2$ and CH$_3$OH in the ice. Therefore, only 1/3 of the CO originally
frozen out will return to the gas phase as CO upon warmup.
}

\maketitle

\section{Introduction}

In the last decade, it has become increasingly clear that the interaction between molecular gas and ice mantles on dust grains in dense clouds
significantly affects the overall chemistry of the cloud as well as the way molecules are observed. The chemistry of a dense cloud is influenced by the presence of ices through reactions taking place on the surfaces of dust grains and by changing the composition of the gas when ice species are desorbed through heating or energetic processing \citep{Aikawa03}. Additionally, molecules may be removed from the gas-phase by freezing out onto dust grains in the early phases of star formation \citep[e.g.][]{Jes02}.  Many observational studies have attested to the high abundance of a large range of molecular ice species in dark clouds and in star forming regions in particular \citep[e.g.][]{Gibb04, Boogert04, Ewine03}. Since interstellar ice can only be observed in absorption toward an infrared source, most of these observational studies have concentrated on ices along single, isolated lines of sight. Large variations in the abundances of some ice species have been observed in different regions, but the lack of spatial information has often made any interpretation difficult. 
Examples of ice species with strongly varying abundances relative to water ice include 
methanol ice \citep{Dartois99, Pontoppidan03a}, OCN$^-$ ice \citep{Pendleton99, Gibb04} and CO ice \citep{Pontoppidan03b}. Other abundant ice species are found to have a fairly constant abundance relative to water ice, such as CO$_2$ \citep{Gerakines99}. This shows that the composition of interstellar ice depends on the physical conditions under which it formed. Thus, for a complete study of the chemical evolution of interstellar ice, spatial information as well as detailed information on the physical environment in which it is found are needed. Infrared spectrometers are now sensitive enough to record spectra along lines of sight toward closely spaced infrared sources, which can be young stars or background stars. This opens the
possibility of directly obtaining spatial information on the distribution of interstellar ices. For example, \cite{Murakawa00} constructed a water ice map of the Taurus molecular cloud using bright background stars. The resulting spatial resolution of this study was $\sim$1 line of sight per 100\,square arcminutes. 
In this article we present high angular resolution (1 line of sight per 0.02 square arcminutes) observations of the distribution of ices in a small region of the Serpens star forming cloud centered on the cluster SVS 4.

SVS 4 is a small but dense cluster of low- to intermediate mass ($L=1-50\,L_{\odot}$) pre-main sequence stars located in the south-eastern core of the Serpens molecular cloud. It was first studied by \cite{EC89}, who found it to be one of the densest YSO clusters known with 11 stars within a region only 12\,000\,AU across, corresponding to a stellar density of $\rm \sim 5\times 10^4\,M_{\odot}\,pc^{-3}$. One of the cluster members, SVS 4-9/EC 95, is the brightest X-ray source in Serpens and is probably a very young intermediate mass star with abnormal coronal activity \citep{Preibisch03}. The center of the cluster is located only $30\arcsec$ or 7500\,AU from the deeply embedded class 0 protostar, SMM 4 \citep[e.g.][]{Hogerheijde99}. Lines of sight toward the stars in the SVS 4 cluster may therefore intersect  part of the outer envelope of SMM 4. This system thus provides a unique opportunity to directly probe the amount of freeze-out near deeply embedded young stellar objects. Detailed millimeter studies of pre-stellar cores \citep[e.g.][]{Caselli99, Bergin02, Tafalla04} and protostars \citep[e.g.][]{Joergensen04} indicate that a significant fraction of the gas-phase molecules
($\gtrsim 90$\% of the condensible species) is depleted onto grains in the cold parts of the envelopes where the timescales for freeze-out are shorter than the age of the core. For the case of protostellar envelopes, J{\o}rgensen et al. propose a `drop' abundance structure where the gas-phase abundances are relatively high in the low density outermost envelope and then drop abruptly to a very low value at the radius where freeze-out becomes effective. In the inner warm regions of the envelope, the ices evaporate resulting in high gas-phase abundances. Our observational ice maps can test such abundance structures directly.

\cite{Pontoppidan03a} presented low-resolution $L$-band spectra toward two of the SVS 4 cluster members, SVS 4-5 and SVS 4-9, and found that the ices in front of these two stars contain a very high abundance of methanol ice of up to 25\% compared to water. This was the first detection of methanol ice outside of high-mass star forming regions and is among the highest abundances observed. Previous searches for methanol ice in low-mass star forming regions have only provided strict upper limits to the abundance of methanol down to a few \% relative to water ice \citep{Chiar96}. The reason for this large variation in the methanol ice abundance is currently not understood. Recent laboratory experiments have shown that methanol can be formed in dense clouds through successive hydrogenation of CO at 10-20\,K on the surfaces of dust grains \citep{Watanabe03}. It is found that the formation of methanol ice along this route depends primarily on the abundance of atomic hydrogen, the thickness of the ice mantle and the temperature of the ice, with the most efficient formation of methanol at 15\,K. Thus, a variation in these parameters could be related to the observed differences in methanol ice abundance. Alternatively, bombardment with energetic protons has also been found to produce moderate yields of methanol in CO:H$_2$O ice mixtures \citep{HM99}. The study of the relation of interstellar methanol ice to the cloud environment in which it is found therefore provides direct constraints on the formation mechanism of this important molecule. 
Due to the high stellar density, the proximity to SMM 4 and the very high methanol ice abundance of the SVS 4 cluster, it presents an excellent case for studying the distribution of methanol-rich interstellar ice on small angular scales in a protostellar environments. 

In this paper we present low-resolution $L$-band spectra of 10 SVS 4 stars obtained with the ISAAC spectrometer on the Very Large Telescope (VLT) at Paranal in Chile.  Additionally, we use archival data from the Infrared Space Observatory (ISO) and the United Kingdom Infrared Telescope (UKIRT) to further constrain the environment. We use the spectra to produce a detailed map of the distribution of water and methanol ice in the SVS 4 region with a spatial resolution of 6\arcsec=1500\,AU. The corresponding line of sight density is one per 0.02\,square arcminutes, which is more than three orders of magnitude higher than previous ice maps.

This paper is organised as follows: \S\,\ref{observations} describes the observations from ISAAC as well as archival data used. \S\,\ref{extSec} shows how the basic observational parameters have been derived.  \S\,\ref{smm4Rel} interprets the
relation to the SMM 4 envelope and \S\,\ref{iceDist} discusses the distribution of ices in the observed region.

\section{Observations and archival data}
\label{observations}
$L$-band spectra of 10 sources in the SVS 4 cluster were obtained using the ISAAC spectrometer mounted on UT 1 of the VLT on the
nights of July 8--9, 2003. The low resolution grating and the 0\farcs6 slit were used, yielding a resolving power of $\lambda/\Delta\lambda\sim600$ and an instantaneous 
wavelength coverage from 2.84--4.15\,$\mu$m. Typical on-source integration times were $\sim$40 minutes per pointing. Additionally,
the two brightest sources, SVS 4-5 and SVS 4-9, were observed in one setting centered on the CH-stretching mode  of solid methanol at 3.53\,
$\mu$m using the medium resolution grating with a resolving power of $\lambda/\Delta\lambda\sim3300$. The spectra were reduced using IDL following the procedure described in \cite{Pontoppidan03b}. The spectra were ratioed by the early-type standard stars, BS 6629 (A0V) and BS 7236 (B9V) to remove telluric absorption lines. 

A 4.07\,$\mu$m image was obtained by co-adding 8 acquisition images from the spectroscopic observations and was flux calibrated relative to the 
acquisition image of the standard star BS 7348 (B8V). The 4\,$\mu$m flux of the standard star was estimated by extrapolating the $V$-band magnitude
of 3.95 to 4\,$\mu$m using blackbody colours, yielding $M_{4\,\mu\rm m}=4.15$.  The spectra were then scaled to the 4.07\,$\mu$m photometric points. 
We estimate the accuracy of the photometry to be better than 0.1\,mag. Standard stars of spectral type A or B are known to have photospheric hydrogen lines in absorption. This may result in excess line
emission in the ratioed spectra. The two standard stars show the presence of the Br$\alpha$ line at the 7\% level relative to the continuum while other hydrogen lines have strengths of $<5\%$. The Br$\alpha$ line was not detected in any of the SVS 4 spectra and any residual emission lines are due to the standard stars. Since the hydrogen lines do not affect the ice bands, no attempt to correct for them was made. 

$J$- $H$ and $K$-band imaging from UFTI at the United Kingdom Infrared Telescope (UKIRT) were extracted from the UKIRT archive, and
ISOCAM 6.7 and 14.3\,$\mu$m imaging as well as a CVF field centered on SVS 4 were extracted from the ISO archive. The near-infrared images were flux calibrated using bright 2MASS stars in the field. Standard aperture photometry was performed using $1.5\arcsec$ apertures. The absolute photometry was found to be better than 0.1 mag when comparing to 2MASS. The ISOCAM data were reduced using the CIA reduction package version DEC01. Because the SVS 4 sources are partially blended at the ISOCAM resolution, the raster maps were resampled to a finer sampling by a factor of two. ISOCAM fluxes were then derived from the broad-band images by fitting a PSF derived from an isolated source in the field to source positions taken from the 4.07\,$\mu$m ISAAC image. The ISOCAM images where then flux calibrated using conversion factors of 2.32\,$\rm adu\,g^{-1}\,s^{-1}\,mJy^{-1}$ and 1.96\,$\rm adu\,g^{-1}\,s^{-1}\,mJy^{-1}$ for LW2 and LW3, respectively \citep{Blommaert00}, 
where adu is the Analog to Digital Unit, and g is the gain. Any colour corrections
to the broad band ISOCAM fluxes were found to be less than 2\%. CVF spectra were extracted using $3\,{\rm pixels}\times 3\,{\rm pixels}=18\arcsec \times18\arcsec$ apertures.

\section{Derivation of observed physical parameters}
\label{extSec}
\subsection{Distance to the Serpens molecular cloud}
There is some controversy in the literature regarding the distance to the Serpens cloud. Distance estimates based on accurate photometry of a few stars ($\sim$10) range from 250 pc \citep{Chavarria88} to 400 pc \citep{ChiarThesis,Hogerheijde99}. However, accurate photometric surveys of larger samples of stars seem to converge on the smaller distance of $\sim 250$\,pc. \cite{Straizys96} find a distance of $259\pm37$\,pc using 105 stars. This result is supported by a preliminary distance estimate of
220-270\,pc using 2MASS photometry of a similarly sized sample (Knude et al., in prep). In this paper all distance-dependent quantities are scaled to a distance of 250\,pc.

\subsection{Determination of extinction and water band continuum}
\label{extDet}
In order to study the distribution of ices toward the SVS 4 cluster, accurate optical depths of the water bands as well as column densities of the refractory dust component must be determined. The optical depths of the water bands depend
sensitively on the continuum chosen. However, since the blue wing of the water band is outside of the atmospheric $L$-band window, the determination of a continuum will always have some degree of uncertainty. One solution is to use a $K$-band spectrum
to define an empirical continuum \citep[e.g.][]{Murakawa00}. Unfortunately, only photometric $JHK$ points are available for the SVS 4 stars. However, as will be shown in this section, most of the sources in SVS 4 have almost no infrared excess. This allows reasonably accurate modeling of the continua using Kurucz stellar atmosphere models. Furthermore, a simultaneous determination of the extinction toward each source can be derived with confidence. 

Since the effective temperatures of low-mass pre-main sequence stars are quite
insensitive to their masses \citep{Siess2000}, Kurucz models with a low $\log g$ and $T_{\rm eff} = 3500-4500$\,K can be used as a rough model of the photospheric emission. In the case of SVS 4-9 and SVS 4-10, the effective temperatures have been determined
spectroscopically by \cite{Preibisch1999}. Any infrared excess is modeled by a single blackbody of $T=500-1500$\,K or by a disk model from \cite{Dullemond01} if the excess is large. The observed SED of each source was de-reddened to fit a Kurucz model
at the $J$, $H$ and $K$ points and a blackbody or disk model was added to fit any infrared excess at longer wavelengths as well as the 
3.8-4.15\,$\mu$m slope of the $L$-band spectra. In cases where a strong infrared excess would also affect the near-infrared points,
the extinction and infrared excess contribution were iteratively corrected until a fit was found, which reproduced both the photometric points and the slope of the $L$ band spectra. This approach was chosen to provide a realistic continuum at 3\,$\mu$m rather than to yield an unambiguous excess model. 
The fits to sources with a strong excess are therefore degenerate to some extent. However, due to the very high absolute values for the extinction ($A_J>5$) the determination of water ice optical depths and extinction was found to be very insensitive within the allowed parameter space, and the uncertainties given on the derived quantities reflect the fitting degeneracy.

\begin{figure}
  \includegraphics[width=8cm]{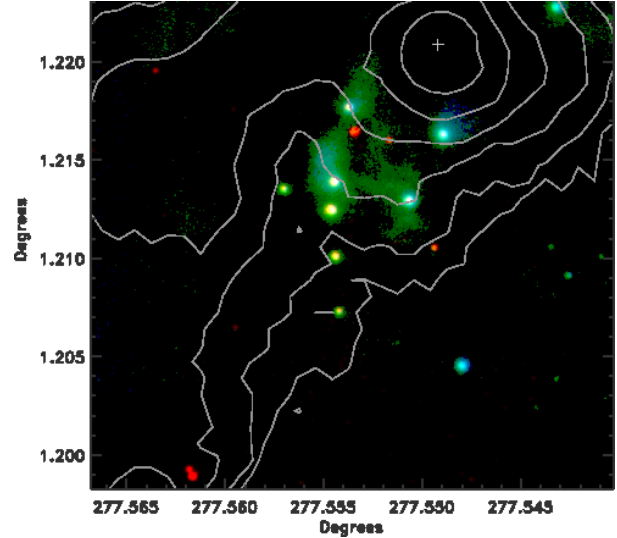}
  \caption{$J,H,4\,\mu$m colour composite of SVS 4. The contours show the 850\,$\mu$m SCUBA map by \cite{Davis99}. The cross indicates the
  central position of SMM 4.}
  \label{SMM}
\end{figure}
\begin{figure}
  \includegraphics[width=8cm]{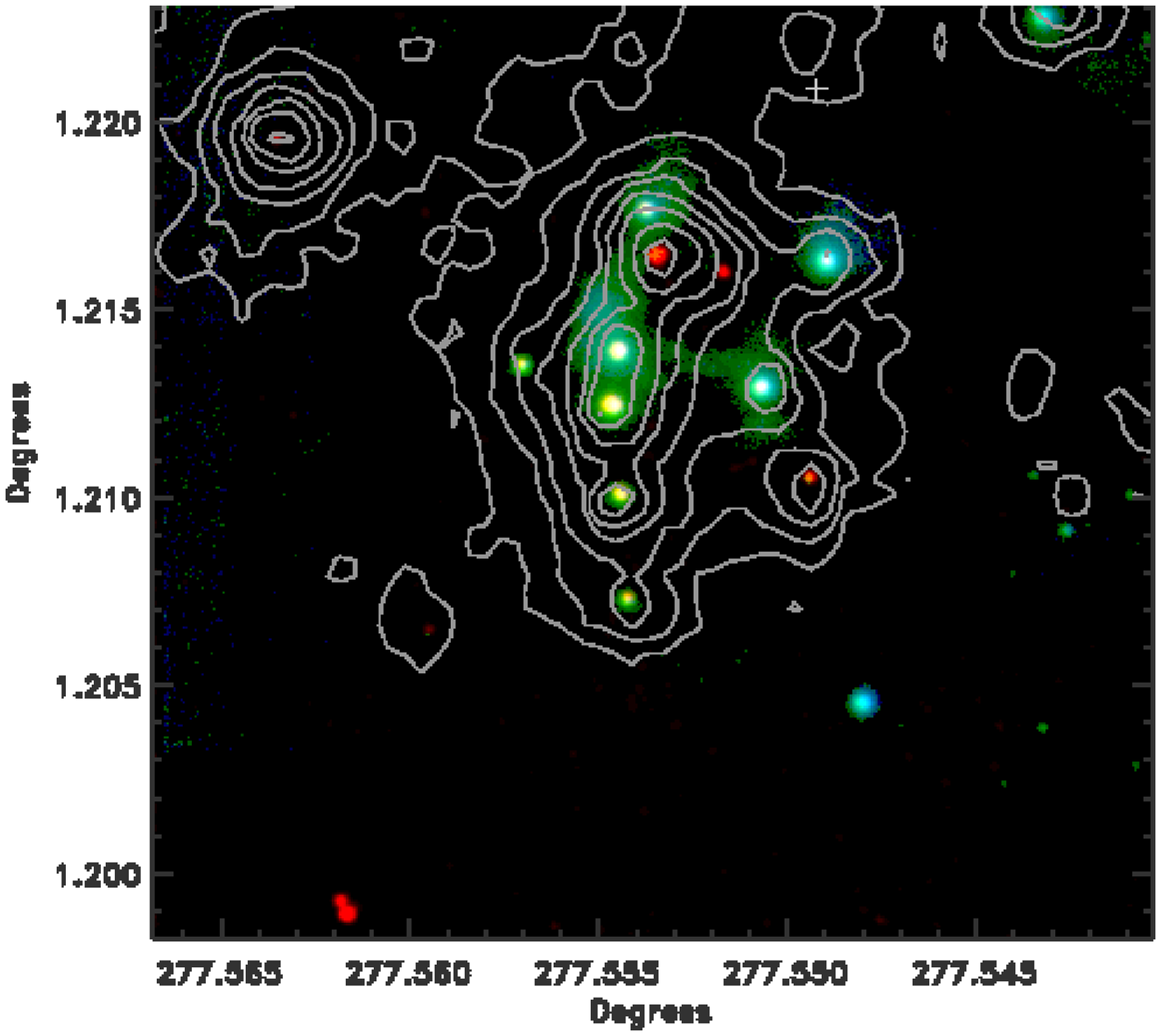}
  \caption{The SVS 4 $J,H,4\,\mu$m colour composite overlaid with ISOCAM 6.7\,$\mu$m contours.}
  \label{6mu}
\end{figure}
\begin{figure}
  \includegraphics[width=8cm]{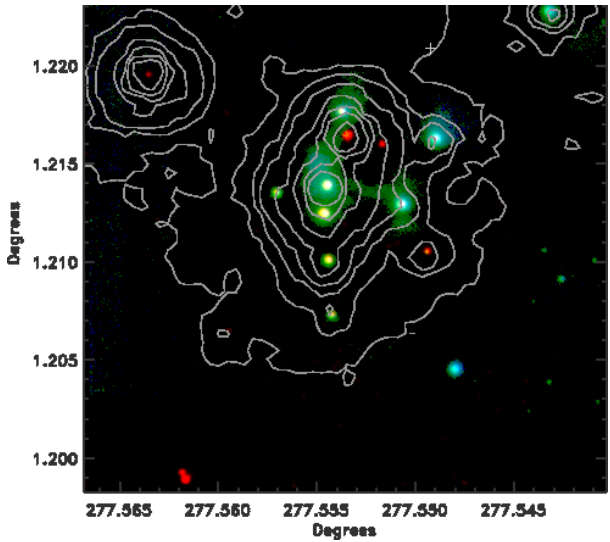}
  \caption{The SVS 4 $J,H,4\,\mu$m colour composite overlaid with ISOCAM 14.7\,$\mu$m contours.}
  \label{14mu}
\end{figure}

The extinction law at near-infrared wavelengths is well described by a power law, $A_{\lambda}/A_{J}=(\lambda/1.25\,\mu {\rm m})^{\alpha}$, where $\alpha=1.6-2.0$ \citep{Cardelli89, Martin90}.  A high value of $\alpha$ has been found to best describe dust in some dense molecular clouds \citep[e.g.][]{Martin90, Lutz96}.  A low value for the power-law index of 1.6 was attempted, but this was found to give extinction-corrected 2--5 $\mu$m fluxes, which were too steep to correspond to a blackbody or Kurucz model for stars with no infrared excess when the $(J-H)$ colours were used to determine the extinction. In particular, an extinction law index of at least 1.85 was required to simultaneously fit all spectral and photometric points of sources like
SVS 4-3, SVS 4-5 and SVS 4-9. A higher index could also be used since this would simply create a small infrared excess. Here, we adopt the minimum value of $\alpha = 1.85$. In the mid-infrared range, the empirical extinction law toward the galactic center by \cite{Lutz96} is used. This is consistent with the ISOCAM extinction law for Serpens ($A_{\rm LW2}=0.41A_K$ and $A_{\rm LW3}=0.36A_K$)  derived by \cite{Kaas04}. The resulting SEDs and water bands are shown in Figs. \ref{SEDs}, \ref{SEDs2} and \ref{SEDs3}. The corresponding fitting parameters are shown in Table 
\ref{PhysPar}. The extinctions are given in terms of the extinction in the $J$-band, $A_J$, to avoid making unnecessary assumptions about shape of the extinction law in the optical wavebands by extrapolating to $A_V$.
 
It is seen that the extinction toward SVS 4 is in general very high ($A_J=5-25$), roughly corresponding an $A_V$ of up to 100\,mag. Most of the sources exhibit very little infrared excess shortwards of 6\,$\mu$m with the exceptions of SVS 4-6 and SVS 4-10. This results in a fairly accurate absolute determination of $A_J$ with an uncertainty dominated by the chosen extinction law
rather than the accuracy of the photometry or, in most cases, the SED model assumptions. We estimate a relative uncertainty
on $A_J$ of 5\%. A different extinction law may cause systematic effects larger than this. The bolometric luminosities of the stars can then be estimated by integrating over the fitted Kurucz models. The derived stellar parameters
are included in Table \ref{PhysPar}. It should be cautioned that the effective temperatures of the stars are not well constrained using photometric
points and should ideally be determined by high resolution near-infrared spectroscopy. The derived luminosities thus depend sensitively on the
assumption that the stars are indeed pre-main sequence stars. As a consistency check the stellar ages and masses have been derived using the evolutionary tracks of \cite{Siess2000}. These are also given in Table \ref{PhysPar}, but are associated with large uncertainties except in the cases
of SVS 4-9 and SVS 4-10, which have spectroscopically determined spectral types.

In order to convert optical depth and extinction to column density, the relations $N_{\rm H_2O}/\tau_{\rm H_2O}=1.56\times 10^{18}\,\rm cm^{-2}$ \citep{Pontoppidan03a} and $N_{\rm H}/A_J = 5.6\times 10^{21}\rm\, cm^{-2}\, mag^{-1}$ \citep{Vuong03} are used. The water ice column density conversion uses a band strength of $2\times 10^{-16}\,\rm cm\,molec^{-1}$ \citep{Gerakines95}.

\begin{table*}
\centering
\begin{flushleft}
\caption{Flux densities and astrometry of SVS 4 sources}
\begin{tabular}{lllllllll}
\hline 
\hline 
Source & RA (J2000) & DEC (J2000) & $1.250\,\mu$m & $1.635\,\mu$m & $2.159\,\mu$m& 4\,$\rm \mu m$ & 6.7\,$\mu$m& 14.3\,$\mu$m\\
 \hline	
SVS 4-2&18:29:56.58&+01:12:59.4	&	$1.8\pm 0.2$	&	$8.5\pm0.9$	&$16.4\pm 1.0$	&$19\pm 2$ 	&$10\pm2$	&$19\pm4$\\
SVS 4-3&18:29:56.70&+01:12:39.0	&	--			&	$0.4\pm0.1$	&$1.6\pm0.2$		&$28\pm 3$&$16\pm4$	&$17\pm3$\\
SVS 4-4&18:29:57.00&+01:12:47.6	&	$1.5\pm0.2$	&	$10.5\pm1$	&$24.5\pm2$		&$35 \pm 4$	&$15\pm4$	&$15\pm3$\\
SVS 4-5&18:29:57.63&+01:13:00.2	&	--			&	$0.03\pm0.01$	&$2.9\pm0.3$		&$130\pm10$ 	&$242\pm50$	&$494\pm100$\\
SVS 4-6&18:29:57.69&+01:13:04.4	&	$0.5\pm0.1$	&	$3\pm0.3$	&$10.3\pm1.0$		&$38\pm4$	&$40\pm6$	&$84\pm20$\\
SVS 4-7&18:29:57.84&+01:12:27.8	&	$0.015\pm0.005$	&	$0.6\pm0.1$	&$4.3\pm0.4$	&$15\pm2$	&$7\pm2$	&$7\pm2$\\
SVS 4-8&18:29:57.88&+01:12:37.7	&	$0.04\pm0.01$		&	$1.8\pm0.2$	&$14.4\pm1$	&$56\pm6$	&$51\pm10$	&$56\pm10$\\
SVS 4-9 (EC 95)&18:29:57.92&+01:12:46.0	&	$0.21\pm0.05$	&	$9.7\pm1.0$	&$68.6\pm7.0$	&$155\pm15$	&$147\pm30$	&$89\pm20$\\
SVS 4-10 (EC 92)&18:29:57.88&+01:12:51.1	&	$0.6\pm0.1$	&	$9.9\pm1.0$	&$41.4\pm4.0$	&$111\pm11$	&$152\pm40$	&$700\pm70$\\
SVS 4-12&18:29:57.22&+01:12:58.4		&	--			&--	&$0.2\pm0.02$				&$26\pm3$	&$24\pm5$	&--\\
\hline
\end{tabular}
\begin{itemize}
\item[] All flux densities are given in units of mJy.
\end{itemize}
\label{Photometry}
\end{flushleft}
\end{table*}

\begin{figure*}
  \includegraphics[width=16cm]{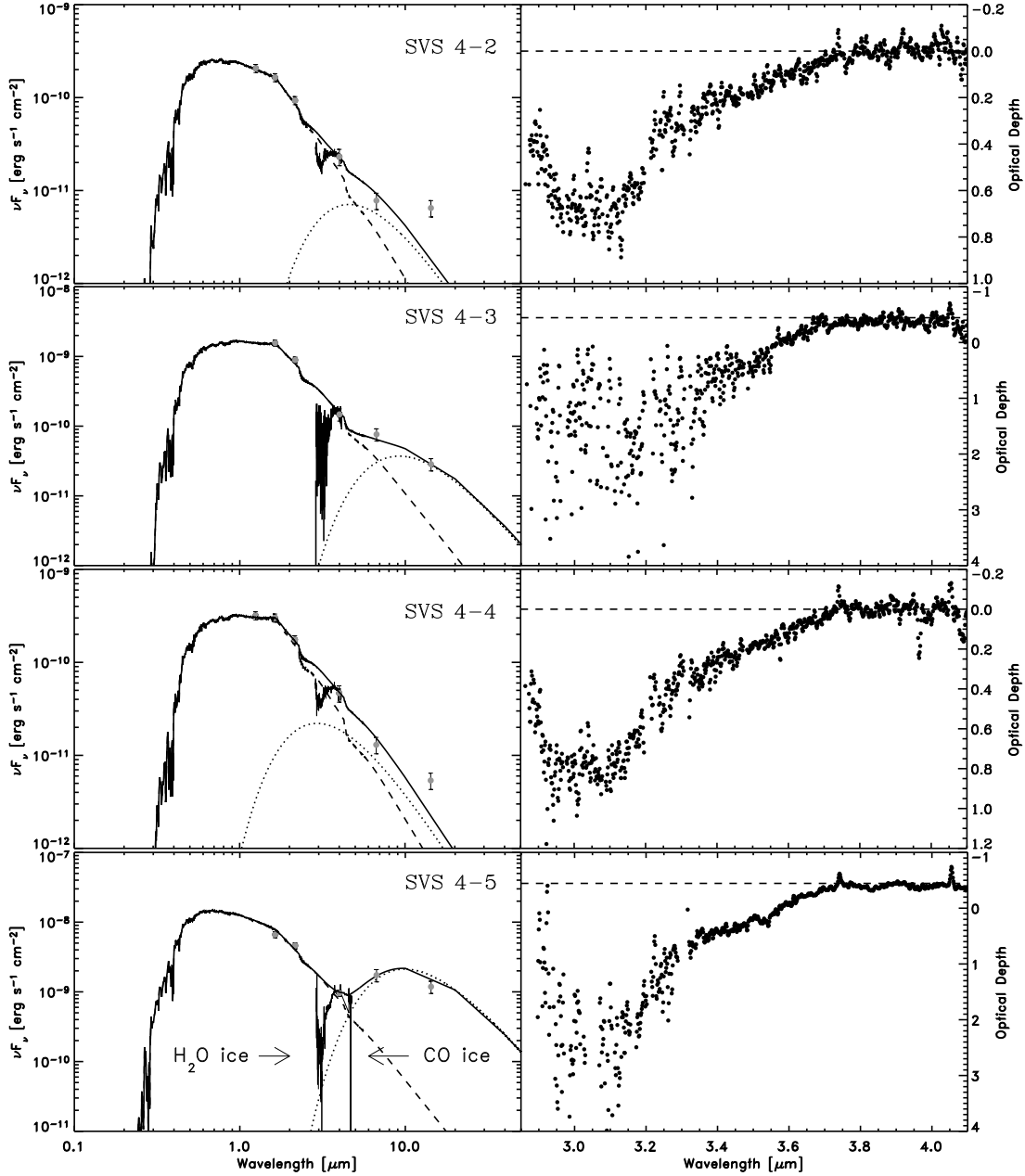}
  \caption{The spectral energy distributions of the SVS 4 stars, which were used to define a continuum for the water bands. The extracted water ice bands are shown next to each SED on an optical depth scale. On the left, the dashed lines show the Kurucz stellar atmosphere model, the dotted line shows the added infrared excess and the full line is the sum of the two contributions.}
  \label{SEDs}
\end{figure*}

\begin{figure*}
  \includegraphics[width=16cm]{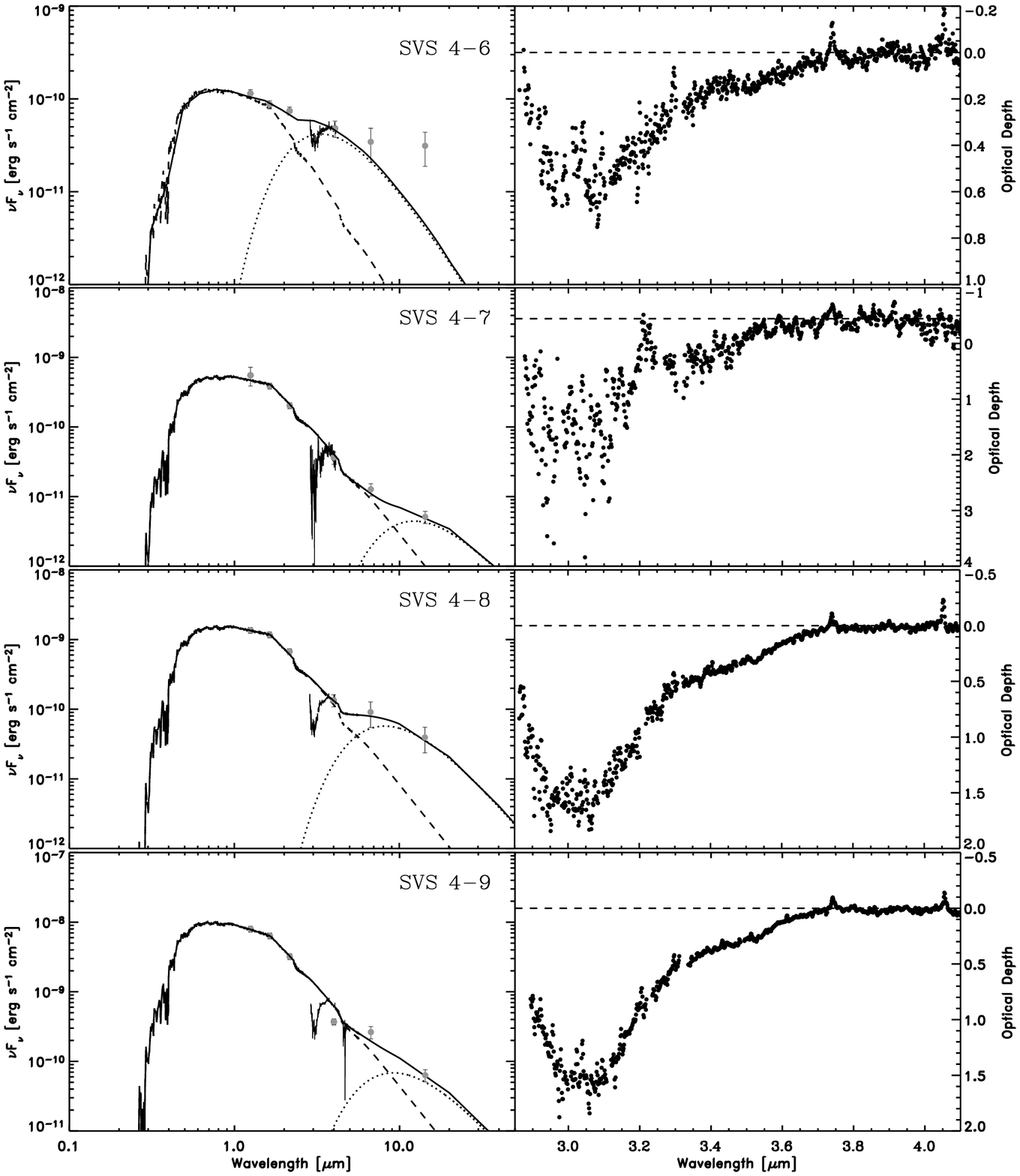}
  \caption{Continued from Fig. \ref{SEDs}.}
  \label{SEDs2}
\end{figure*}

\begin{figure*}
  \includegraphics[width=16cm]{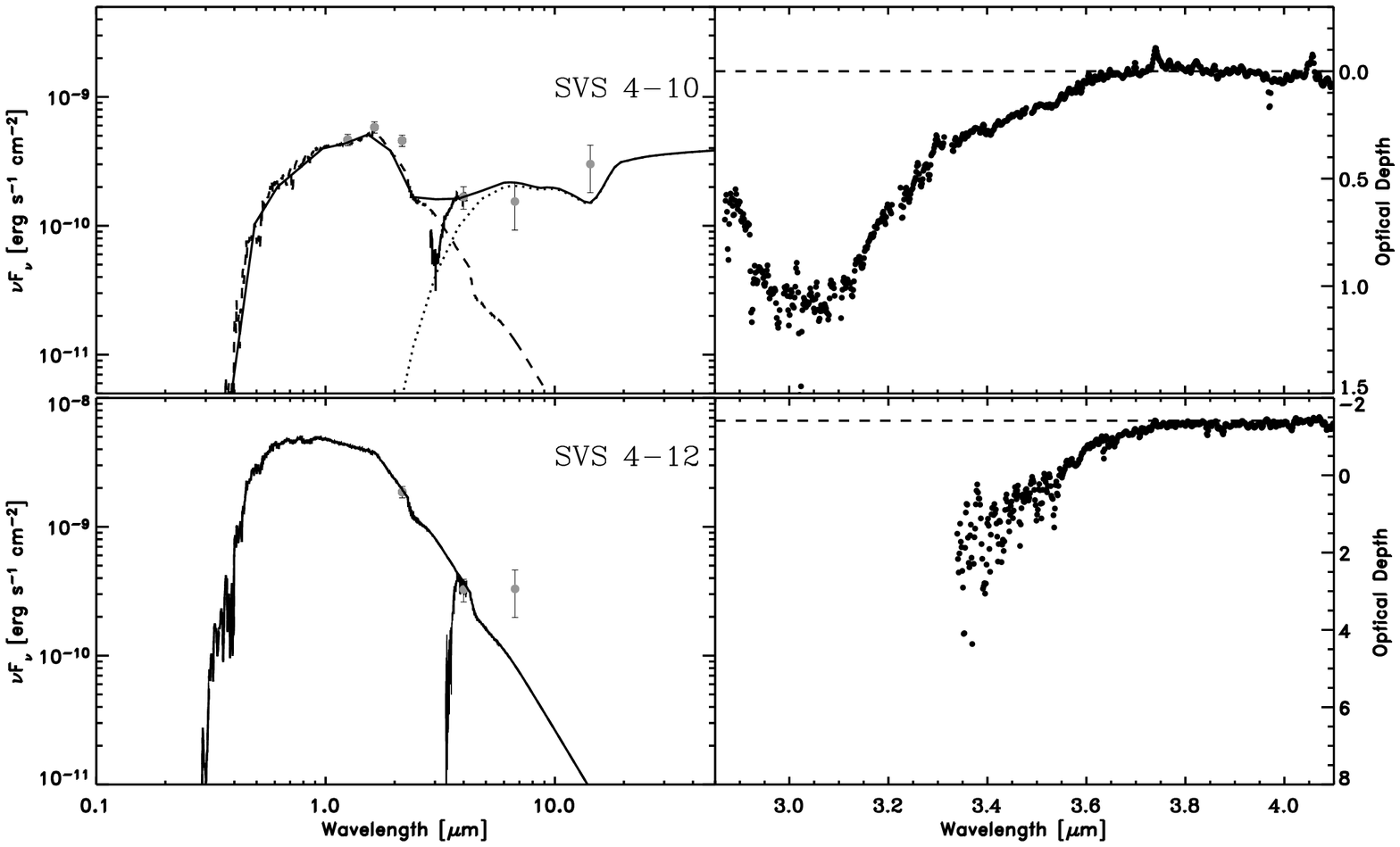}
  \caption{Continued from Fig. \ref{SEDs2}.}
  \label{SEDs3}
\end{figure*}

\begin{table*}
\centering
\begin{flushleft}
\caption{Physical parameters for SVS 4 sources}
\begin{tabular}{llllllllll}
\hline 
\hline 
Source & $L_{\rm bol}$  &$T_{\rm eff}$ & stellar mass     & Age  & $A_J^a$ & $\tau_{\rm H_2O}$ & $N_{\rm H_2O}$             & H$_2$O ice abundance        \\
              & [L$_{\odot}$]   & [K]                   & [M$_{\odot}$]  & [Myr] & [mag]   &                                   &  [$10^{18}$\, cm$^{-2}$]  &  $\times 10^{-5}$ w.r.t. H$_2$ \\
\hline
SVS 4-2          &0.7&4750&1.5&17 &4.2  &$0.7\pm0.1$&$1.1\pm0.2$&$9.3\pm1.4$\\
SVS 4-3$^*$  &4.9&4250&1.0&0.5&18.0&$3.5\pm0.5$&$5.5\pm0.8$&$10.9\pm1.6$\\
SVS 4-4          &0.9&4250&1.0&3   &4.9   &$0.85\pm0.1$&$1.3\pm0.2$&$9.7\pm1.2$\\
SVS 4-5          &38 &4500&3.5&0.3&21.0 &$3.5\pm0.3$&$5.5\pm0.5$&$8.6\pm0.8$\\
SVS 4-6          &0.4&4750&0.8   &42   &4.9  &$0.7\pm0.1$&$1.1\pm0.2$&$9.1\pm1.4$\\
SVS 4-7          &1.5&4500&1.3&3   &10.5&$2.0\pm0.3$&$3.1\pm0.5$&$10.6\pm1.8$\\
SVS 4-8          &4.6&4500&1.4&0.8&10.8&$1.7\pm0.2$&$2.7\pm0.3$&$8.8\pm1.2$\\
SVS 4-9          &27&4750&3.5&0.4&10.5&$1.6\pm0.1$&$2.5\pm0.2$&$8.5\pm0.8$\\
SVS 4-10       &1.2&3750&0.5&0.8&6.2&$1.1\pm0.1$&$1.7\pm0.2$&$9.9\pm1.2$\\
SVS 4-12$^*$ &28&5000&3.5&0.5&26&$8\pm1$&$12.5\pm2.0$&$17.2\pm2.4$\\
\hline
\end{tabular}
\begin{itemize}
\item[$^a$] Estimated relative uncertainty of 5\%.
\item[$^*$] Not detected in $J$, making the derived spectral type and age uncertain.\\
\end{itemize}
\label{PhysPar}
\end{flushleft}
\end{table*}

\subsection{Derivation of the extinction and water ice optical depth toward SVS 4-12}
\label{svs412}
The determination of the extinction and water optical depth toward SVS 4-12 must be discussed in detail, since this source is not detected in the $J$- and $H$- band images from UKIRT and since the water ice band is heavily saturated. Only a $K$-band point, the 3.8-4.2\,$\mu$m spectral slope and an uncertain 6.7\,$\mu$m ISOCAM point are available for the extinction determination.  In principle, the lack of near-infrared photometric points prevents an accurate determination of extinction if the source has a significant infrared excess. However, an extinction of $A_J=26\,$mag gives a good fit to a Rayleigh-Jeans slope of a stellar spectrum, without any infrared excess (see Fig. \ref{SEDs3}). The presence of any excess will result in a smaller extinction, since the dereddened SED in that case will be shallower. Therefore, the derived extinction is a conservative upper limit. However, the simultaneous good fit of the $K$-band point and the slope of the ISAAC spectrum indicates that SVS 4-12 has a naked photospheric SED at least to wavelengths of 4\,$\mu$m, although the current data do not allow us to rule out the possibility that SVS 4-12 has some minor infrared excess at longer wavelengths.

\begin{figure}
  \includegraphics[width=8.5cm]{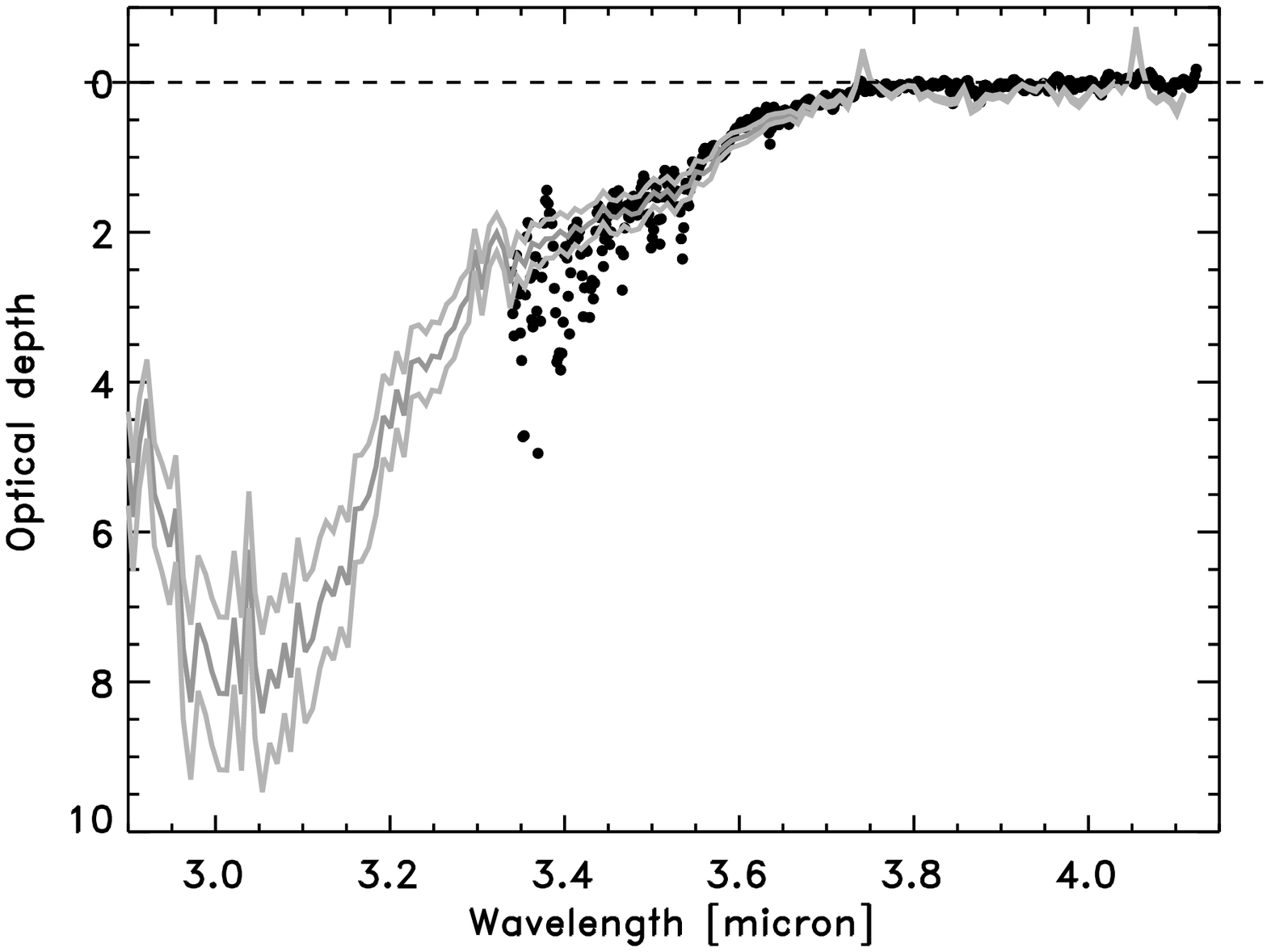}
  \caption{Determination of the optical depth of the water ice band toward SVS 4-12. The dots show the parts of the SVS 4-12 $L$-band spectrum which are detected with more than 5$\sigma$. The curves are the spectrum of SVS 4-9 scaled to optical depths of 7, 8 and 9 to indicate the allowed range of depths of the SVS 4-12 water ice band.}
  \label{ice_comp_svs4-12}
\end{figure}

The optical depth of the water ice band toward SVS 4-12 is determined by scaling the water band from SVS 4-9 to the SVS 4-12 red wing. In general, all the water ice bands observed toward the SVS 4 sources have very similar shapes, i.e. one water band can generally be scaled to fit well with another at the 10\% level. The best
fit of $\tau_{\rm H_2O}=8\pm1$ for SVS 4-12 using the water band of SVS 4-9 is shown in Fig. \ref{ice_comp_svs4-12}. This value does not change significantly if other water bands from SVS 4 are used. Thus if the extinction is overestimated due to an infrared excess in SVS 4-12, the measured abundance of water ice relative to the dust column density will be a strict lower limit. 

\section{Relation to the envelope of SMM 4}
\label{smm4Rel}

The projected distances of the individual SVS 4 stars
to the central submillimeter position of SMM 4 are 15-40\arcsec, corresponding to 3\,750 - 10\,000 AU at a distance of 250 pc. SMM 4 is also associated with a large scale molecular outflow oriented NW-SE. Consequently,  
the projected south-eastern lobe of the outflow passes through the SVS 4 region. Therefore, there is a
possibility that the lines of sight toward some SVS 4 stars may probe the outer envelope of SMM 4 and/or material which has
been processed by shocks created by the outflow. 

It is therefore an important issue to determine the depth of the SVS 4 cluster relative to the envelope and outflow of SMM 4. Clearly, much of SVS 4 share the line of sight with the SMM 4 envelope as seen in Fig. \ref{SMM}. However, it is not clear whether the SVS 4 stars are situated in front of, inside, or behind the envelope of SMM 4. Only the two last scenarios allow for a study of the ices in the envelope of a class 0 protostar since the ice absorption bands do not probe what lies behind the individual SVS 4 stars. The last scenario would require the presence of a second dense cloud in front of SVS 4, which can produce the
ice absorption and extinction corresponding to H$_2$ column densities of up to $7\times 10^{22}\,\rm cm^{-2}$ for SVS 4-5 and SVS 4-12. The Serpens cloud was mapped in a range of
optically thin molecular lines by \cite{Olmi02} who found a column density through the SVS 4 region of $6\times 10^{22}\,\rm cm^{-2}$ within a large beam of $\sim 50\arcsec$. However, this column density was measured using C$^{18}$O, a molecule known to be partially frozen out at low temperatures and high densities. The direct measurement of the CO ice abundance discussed in Sec. \ref{COice} shows that the gas-to-solid ratio of CO is 2-5 toward SVS 4, assuming a total abundance of CO of $2\times 10^{-4}$. However, if the  CH$_3$OH and CO$_2$ ices have been chemically formed from hydrogenation and oxidation of CO molecules, the fraction of CO molecules directly observed to be removed from the gas-phase increases by more than 50\% (see \S\ref{iceDist}). Taking the freeze-out into account, the H$_2$ column densities may increase to $12\times10^{22}\rm\,cm^{-3}$ in the most extreme case. The observed column densities can therefore not completely rule out the presence of a dense cloud containing SVS 4 and the observed ices, which is unrelated to SMM 4. A high resolution map ($\sim 20\arcsec$)
of the C$^{18}$O (2-1) line by \cite{White95} shows that SVS 4 is located on the edge of the molecular ridge of the south-eastern clump in Serpens, but shows no evidence of a separate, extended clump of dense gas with very high column density associated with SVS 4. Additionally, the C$^{18}$O line has a single gaussian shape with a $FWHM$ of $\sim 2\,\rm km\,s^{-1}$ in the direction of SMM 4, indicating that no other large scale velocity component is present. It will be assumed that SVS 4 is not associated with a foreground cloud, although it should be noted that this issue is not fully resolved. 

To decide whether SVS 4 can be placed behind or significantly inside the envelope of SMM 4, a physical model for the envelope was calculated using the procedure described in \cite{Jes02}. The model was fitted to the SCUBA 850\,$\mu$m map by
\cite{Davis99} and the ISO-LWS spectrum extracted by \cite{Larsson00}. The resulting model has a power law density profile with index -1.6 and 
an outer radius of slightly less than 10\,000\,AU where the temperature drops to 9\,K. The total luminosity is 11\,$L_{\odot}$, the density at 1000\,AU is $5\times 10^6\,\rm cm^{-3}$ and the total mass is 5.8\,$M_{\odot}$. These values are largely consistent with earlier simple models by \cite{Hogerheijde99}. 

\begin{figure}
  \includegraphics[width=8cm]{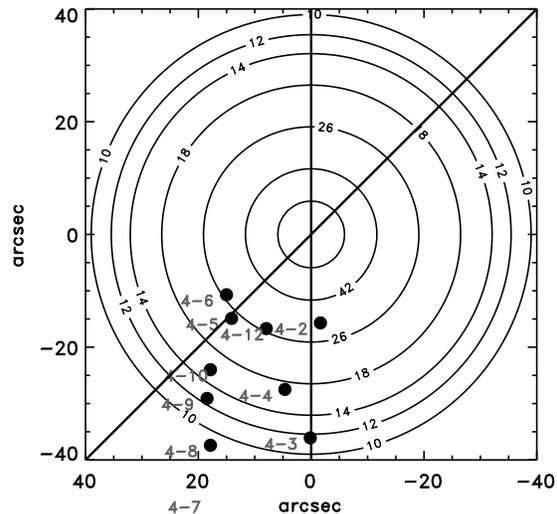}
  \caption{Contours showing the projected extinction in the $J$-band through the model envelope of SMM 4 compared to the positions of the
  SVS 4 lines of sight. The solid lines indicate the range of observed position angles of the
  outflow associated with SMM 4.}
  \label{AjContours}
\end{figure}

The projected extinction through the envelope of SMM 4 calculated from the model density profile is shown in Fig. \ref{AjContours}. Clearly, the extinction through the SMM 4 envelope is too high for all the SVS 4 stars to lie entirely behind the SMM 4 envelope. However, SVS 4-5 and SVS 4-12, which both have measured extinctions of $A_J\sim 25\,$mag, lie very close to the $A_J=26$ contour. This places them effectively behind the envelope if all the extinction is caused by SMM 4. Conversely, SVS 4-2 and SVS 4-6 must in any case lie in the frontal part of the envelope because their measured extinctions are much less than those predicted by the SMM 4 model. It is interesting to note that this behaviour is expected if SVS 4 is a roughly spherical cluster directly embedded into the envelope of SMM 4. In other words, since the angular extent of the cluster is similar to the depth of the envelope ($\sim 10\,000$\,AU), it is expected that some sources lie in front of the envelope and some behind. This geometry is thus consistent with the drastic variation of extinction from source to source on very small angular scales.

\section{The distribution of ice toward SVS 4}
\label{iceDist}
\subsection{Constructing an ice map}
Once the ice column densities and infrared extinctions are determined, maps can be constructed showing the spatial distribution of these 
quantities toward SVS 4. Some caveats apply to such maps. Since only pencil-beam lines of sight are probed, no information about the ice content
of the cloud between the lines of sight can in principle be inferred. However, under the assumption that the observed lines of sight resolve any
variations in ice abundance, a map can be constructed using interpolation techniques. In this work, the IDL 2-dimensional kriging interpolation algorithm is used.

The creation of an ice map is similar to the construction of extinction maps of dark clouds using background stars \citep[e.g.][]{Alves01}. Due to sensitivity limitations, background stars can often not be used to obtain ice maps of high spatial resolution. In this article, a densely packed collection of stars embedded in the molecular cloud rather than background stars are used, and consequently great care has to be taken in interpreting the resulting distribution of ice. One problem concerns the 
interpretation of differences in column densities, since embedded stars are likely to probe different depths through the cloud. This problem will not affect maps of abundances of the ice relative to extinction or other ice species. The second caveat concerns the interaction of the embedded stars with the ice in their immediate surroundings. We will show in the following section that this is unlikely to cause problems for stars of low luminosity ($L\sim 1\,L_{\odot}$). 
Conversely, maps can be used to study the interaction of individual embedded stars with local ice.  

\subsection{Water ice}
\subsubsection{The water ice map}
\begin{figure}
  \includegraphics[width=8.5cm]{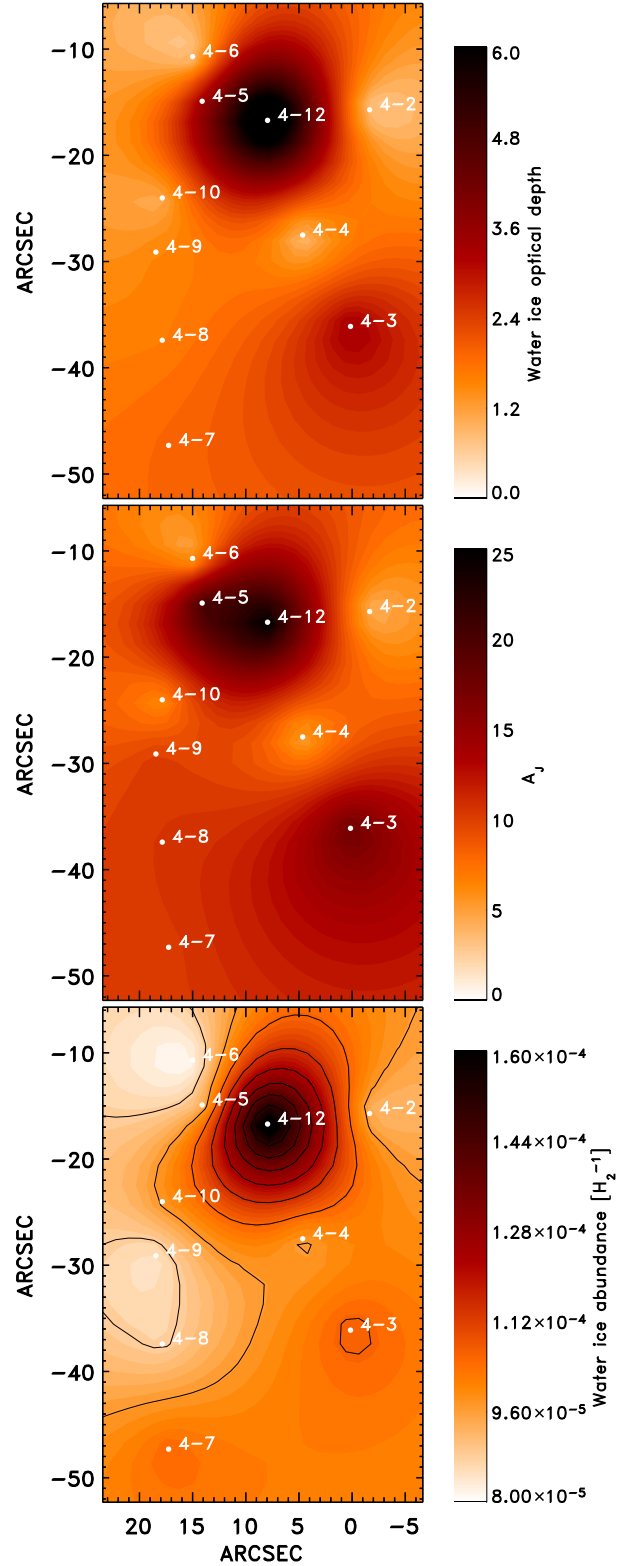}
  \caption{Top: Map of the water ice optical depth toward SVS 4. Middle: Map of $A_J$ of the same region. Bottom: Map of the abundance of water ice relative to H$_2$. Filled circles indicate the positions of the measured sight-lines. SMM 4 is centered on (0,0). }
  \label{waterPlot}
\end{figure}

Maps of the water ice column density, extinction and water ice abundance toward SVS 4 are shown in Fig. \ref{waterPlot}. The water ice abundance map is simply the
ratio between the water ice column density map and the extinction map. It thus shows the line of sight averaged abundance of water ice relative to the refractory dust component or H$_2$ column density in front of the SVS 4 stars.  It can be seen that the extinction and the column density
of water ice vary with factors of 5 and 10, respectively, on angular scales of only $\sim10\arcsec$. The largest extinction and water ice column density are observed toward SVS 4-12. The abundance of water ice relative to
molecular hydrogen is also seen to peak sharply toward SVS 4-12 by a factor of 1.9. Within the framework of the envelope model of SMM 4, the line of sight toward SVS 4-12 passes through a gas density of $4\times 10^{5}\,\rm cm^{-3}$ and
a temperature of 11\,K. Furthermore, the abundance rises slightly in the south-western part of the cluster, parallel to the dense ridge in the south-eastern Serpens core seen in molecular gas \citep{White95}.

\begin{figure}
  \includegraphics[width=8cm]{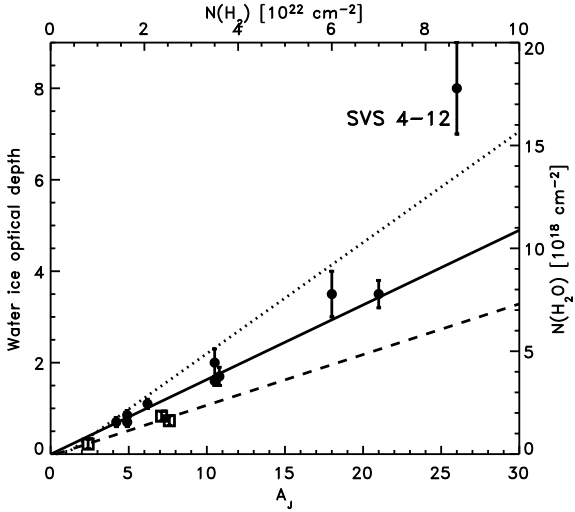}
  \caption{Relation between extinction and water ice column density for the SVS 4 sources (filled circles) and
  the 3 unrelated Serpens YSOs, EC 90A+B and EC 82  (open squares). The solid line indicates the best-fitting line through the SVS 4 
  points, excluding SVS 4-12. The dashed line is the best-fitting line through the three comparison Serpens sources and the dotted line is the
  relation measured for the Taurus cloud by \cite{Whittet01} (see text).}
  \label{AVvsTau}
\end{figure}

The relation between the infrared extinction and the water column density is show in Fig. \ref{AVvsTau}. The derivative of this relation basically measures the water ice abundance as a function of depth into the cloud. Note the important distinction to the absolute line of sight abundance, which is simply the ratio of ice column density to H$_2$ column density. The derivative of the abundance relation is a measure of the {\it local} ice abundance averaged over the angular dimension of the observed cloud. An absolute abundance
is an average over local abundances and does for example not remove any contribution from bare grains along the line of sight. The absolute ice abundance can thus be seen as a lower limit to the local ice abundance.

The lines of sight toward the SVS 4 sources exhibit a linear relation or constant water ice abundance between $A_J=5-25$ except for the line of sight toward SVS 4-12, which is over-abundant in water ice by a factor of 1.9 relative to the other SVS 4 sources. The best-fitting least-squares line for SVS 4 between $A_J=4-25$ and excluding SVS 4-12 is:
\begin{equation}
\tau_{\rm H_2O} =(0.07\pm0.09)+(0.15\pm0.01)\times A_J
\end{equation}
The slope of the derived relation corresponds to a local abundance of water ice relative to $\rm H_2$ of 
$(9\pm1)\times10^{-5}$. Extrapolating the linear relation to small extinctions yields a negative intercept with the $A_J$ axis of $-0.5\pm0.6$. 
However, the negative intercept is 
not statistically significant and the relation could also pass through the origin. Note that a negative intercept with the $A_J$ axis, if significant, translates into the presence of water ice without a refractory dust component. This is an unlikely scenario, and more plausible explanations of a negative threshold would be that a systematic error plays a role or that the relation is no longer linear for small $A_J$.  Significantly, the SVS 4 sources seem to exclude a positive threshold for the water ice in the $\tau_{\rm H_2O}-A_J$-relation of $A_J\sim 1$ such as found in general for molecular clouds \cite[e.g.][]{Whittet88} and for Serpens in particular \citep{EH89}. The two sources SVS 4-3 and SVS 4-7
have a slightly, but significantly, higher water ice abundance compared to the best fitting line of about 20\%. These two sources are responsible for the
higher water abundance in the south-western part of SVS 4 as seen in Fig. \ref{waterPlot}. 

For comparison, lines of sight toward three other young stellar sources from the south-eastern clump in Serpens (EC 90A+B and EC82), unrelated to SVS 4, have also been plotted in Fig. \ref{AVvsTau}. These objects were observed and analysed using the same approach as for the SVS 4 sources. It is found that the other Serpens sources are systematically and significantly under-abundant in water ice compared to SVS 4 and result in a relation with a small, but positive intercept with the $A_J$ axis of $A_J = 0.4\pm1.3$. This is roughly consistent with previous determinations of the water ice--extinction relation for Serpens \citep{EH89}, but also consistent with the relation passing through the origin. This may be an indication of a higher degree of ice evaporation near these sources or a lower ice formation efficiency outside the densest parts of the
Serpens core. 

The corresponding relation for the Taurus molecular cloud of $\tau_{\rm H_2O}=0.072\times (A_V-3.2)$ is included in Fig. \ref{AVvsTau} \citep{Whittet01}.
The dense cloud extinction law derived by these authors in Taurus, $A_V=rE_{J-K}$ with $r=5.3$,
was used to convert the optical extinction to $A_J$, giving $\tau_{\rm H_2O}=0.24\times(A_J+0.95)$. The
conversion is relevant for extinctions above the ice threshold. The water ice relation for Taurus
was determined using sources with $A_V < 26$, roughly corresponding to $A_J\lesssim 7.5$. It is seen to be
consistent with the SVS 4 points out to $A_J\sim 6$, but deviates significantly from the best-fitting line through the SVS 4 points. According to the Taurus relation, the water ice abundance in Taurus is $1.35\times 10^{-4}$. In other words, assuming that the grain size distribution in SVS 4 and in Taurus are identical, the ice mantle volume must be on average 40\% smaller in SVS 4 compared to Taurus in order to create the SVS 4 ice abundance. However, the abundance of water ice toward SVS 4-12 is significantly higher than that of the Taurus cloud.  

\subsubsection{A jump in the abundance of water ice in the SMM 4 envelope?}

\begin{figure}
  \includegraphics[width=8.5cm]{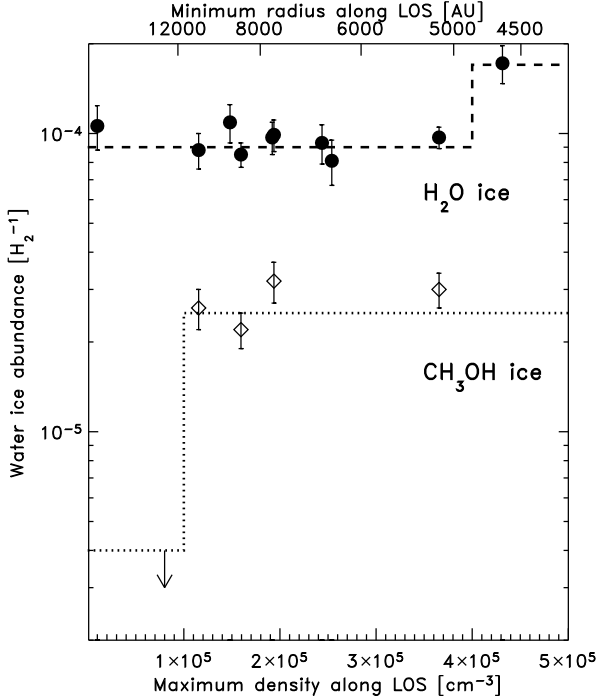}
  \caption{Relation between the water and methanol ice abundances and the SMM 4 envelope model. The bottom axis indicates the maximum density
  encountered by each line of sight. The top axis indicates the corresponding minimum distance to the center of the SMM 4 envelope.
  The dashed line sketches the sharp jump in water ice abundance around 5000\,AU from the center of SMM 4.
  The dotted line sketches the jump in methanol ice abundance observed at a radius of $\sim$12\,000\,AU
  relative to the center of SMM 4. Outside this radius only strict upper limits on the abundance of solid methanol are available (see Sec. \ref{methIce}).}
  \label{JumpPlot}
\end{figure}

The very high abundance of water ice observed toward SVS 4-12 merits further scrutiny. Fig. \ref{JumpPlot} shows the observed water abundance toward
each SVS 4 source as a function of the maximum density (or the minimum distance to SMM 4) probed by each line of sight. The densities and distances were
calculated using the model profile of SMM 4 and the measured extinctions of the individual sources to estimate their depth in the SMM 4 envelope. SVS 4-12
is the source which probes the deepest into the envelope at a minimum distance to the center of SMM 4 of 4700\,AU and a maximum density of $\rm 4.3\times 10^{5}\,cm^{-3}$. In the figure, a sharp jump in the water ice abundance is evident just below 5000\,AU, while the abundance remains constant at larger distances. 
However, the presence of the jump depends on only one source, SVS 4-12. As discussed in Sec. \ref{svs412}, we regard the abundance measurement toward SVS 4-12 relative to those of the other SVS 4 sources as robust or possibly a lower limit. As discussed by \cite{Joergensen04}, the amount of freeze-out is expected to increase significantly at the envelope radius where the
density is high enough for the freeze-out timescale to become shorter
than the age of the core. For a typical lifetime of $\sim 10^5$ yr,
this is expected to occur at densities of a few $\times 10^5$\,cm$^{-3}$,
comparable to the density at which the jump occurs. Of particular interest is
the apparent sharpness of the jump, which seems to be no more than 500 AU
wide. If real, this provides strong constraints on such models
and justifies further ice mapping of protostellar envelopes. Unfortunately, the data presented here only give constraints on the abundance of water ice toward SVS 4-12 and additional deep mid-infrared spectroscopy is required to study the abundance jump in other ice species.

\subsubsection{Caveats}
To test the conclusions reached by analysing the $\tau_{\rm H2O}-A_J$-relation, sources
of systematic errors should be explored. For example, a shallower infrared extinction law would result in a systematic underestimate of the extinction of up to 10\%, sufficient to cause the relation to pass through the origin. However, as discussed in Sec. \ref{extDet}, a small index for the extinction law is not favoured. Also, this effect would influence the three comparison sources. The errors in the individual water optical depths are often larger than 10\%, but systematic effects are estimated to be smaller. The dominant systematic sources of error in the optical depths are related to the adopted model
for the intrinsic SED. This error can be estimated by comparing the shape of the water ice band for sources with and without a large infrared excess.
In Figs. \ref{SEDs}, \ref{SEDs2} and \ref{SEDs3} it is seen that all water ice bands have similar shapes regardless of the shape of the background source SED. This is a clear indication that the derived optical depth is reasonably model independent. 

Finally, it can be expected that the stars heat the
dust in their immediate vicinity to temperatures high enough to evaporate water ice. The question is if this dusty, but ice-free region is large enough to significantly affect the observations. A zone of dust with no ice mantles around each star would create a threshold in the $\tau_{\rm H_2O}-A_J$-relation.
We used a 1-dimensional radiative transfer model of a 50\,$L_{\odot}$ star embedded in a cloud of constant density, calculated using the DUSTY code \citep{DUSTY}, to estimate the threshold created by the radiation of the star. These models show that the temperature rises above an assumed water sublimation temperature of 90\,K only within a radius of 300\,AU. For a star of 1\,$L_{\odot}$, the sublimation temperature is reached at 50\,AU. The contribution to the extinction
from bare grains is then $A_J\simeq 0.7\,$mag for the most luminous stars in SVS 4 and $A_J\simeq 0.1$ for the least luminous stars. Correcting
for this contribution using the calculated luminosity of each star brings the $\tau_{\rm H_2O}-A_J$-relation slightly closer to the Taurus relation. 

No line of sight is expected to pass through the ice evaporation zones from any of the other stars. If the density around the stars is increasing inwards, i.e., if each star in SVS 4 is associated with a remnant envelope, the column density of warm, bare grains and thus the ice threshold will increase. Conversely, this scenario is unlikely since no significant ice threshold has been observed. A full map of CO ice toward the SVS 4 stars will be much more sensitive to moderate heating of the envelope material due to the higher volatility of CO ice. The limited CO ice observations toward SVS 4 sources described in Sec. \ref{COice} do indicate that a significant fraction of the dust in the region presently has temperatures of $\sim20$\,K, in agreement with moderate heating of the cloud and no significant density gradients near the SVS 4 stars.

\subsection{CO ice}
\label{COice}
An additional constraint on the temperature of the ice can be found from high resolution spectroscopy of CO ice in the stretching vibration mode around $2139\,{\rm cm^{-1}} = 4.67\,\mu$m. Laboratory experiments by \cite{Collings03} indicate that a layer of pure CO ice will partly migrate into an underlying layer of porous water ice upon warmup. The different types of CO ice can be distinguished by the shapes of their corresponding spectral profiles \citep{Tielens91}. \cite{Pontoppidan03b} suggested that the relative column densities of the pure CO ice band at 2139.9\,$\rm cm^{-1}$ and of CO ice 
trapped in water ice at $\rm 2136.5\,cm^{-1}$ are a sensitive indicator of the thermal history as well as current temperature of interstellar ice due to the migration effect. In this scenario, a strong band of pure CO ice indicates the presence of ice with temperatures less than $\sim 20\,$K, while a strong band of water-rich CO indicates that a significant amount of ice at some point in time has been heated to temperatures higher than 20\,K. 

Three of the SVS 4 stars have been observed in CO ice (see Table \ref{COiceTab}). The spectrum of SVS 4-10 by \cite{Chiar94} is not of high enough quality to extract quantitative information about column densities of different CO ice components, although a rough estimate is given in Table \ref{COiceTab}. The observed total line of sight column densities of CO ice in SVS 4-5 and SVS 4-9 are $2.7\times 10^{18}\,\rm cm^{-2}$ and $2.5\times 10^{18}\,\rm cm^{-2}$, respectively. However, since SVS 4-5 has an extinction in the $J$-band of 22.8 magnitudes, while SVS 4-9 has an extinction of 10.5, the total line of sight CO abundance for SVS 4-9 ($8.5\times 10^{-5}$) is twice that of SVS 4-5. Interestingly, 
the line of sight abundances of pure CO toward SVS 4-5 and 9 are almost identical at $3\times 10^{-5}$, while a factor of almost four distinguishes the abundances of CO in water. The estimated
column density of pure CO toward SVS 4-10 from \cite{Chiar94} is consistent with a constant local abundance of pure CO in SVS 4. 

Although it is not possible to identify any unambiguous relationship using only two points, some interesting suggestions can be made. Clearly, the amount of
water-embedded CO ice on a grain varies significantly through the cloud. This stands in sharp contrast to the pure CO ice, which seems
to have a constant local abundance, based on three sources. This is difficult to explain from simple migration of CO from the pure form to the water-embedded form upon warmup, but also does not conform to a scenario where the water-embedded CO has formed together with the water mantle. In the case of efficient migration, a large component
of CO in water should be reflected by a correspondingly smaller component of pure CO. In the simultaneous formation case, the water-embedded CO might be expected to show a constant local abundance. Instead, the CO ice abundances may point to a more complex thermal history of the cloud material. A constant local
abundance of pure CO might indicate recent freeze-out, while the abundances of water-embedded CO could indicate thermal processing of an earlier
CO layer. Also, a processing scenario other than heating which drastically changes the water-rich CO abundance could easily change the abundances of other ice species, such as methanol and CO$_2$.
This is not observed to the same degree as for CO (see Secs. \ref{methIce} and \ref{CO2ice}) In this picture, the early thermal processing might have been caused by the formation of the SVS 4 cluster, while the recent freeze-out of CO might be related to the formation of SMM 4. Alternatively, 
the CO ice bands may show processing from interaction of the SMM 4 outflow with the envelope material. A complete map of SVS 4 in CO ice will be needed to explore this question further. 

\begin{table*}
\centering
\begin{flushleft}
\caption{CO ice observations toward SVS 4 sources}
\begin{tabular}{llllll}
\hline 
\hline 
Source & N(Pure CO) & N(CO in water) & Abundance (pure CO) & Abundance (CO in water) & Reference\\
              &[$10^{18}\rm cm^{-2}$]&[$10^{18}\rm cm^{-2}$]&[$10^{-5}$] w.r.t. H$_2$&[$10^{-5}$] w.r.t. H$_2$&\\
\hline
SVS 4-5 &$1.8\pm 0.4$&$0.9\pm0.1$&$2.8\pm0.7$&$1.4\pm0.2$&a\\
SVS 4-9 &$0.9\pm 0.15$&$1.6\pm0.1$&$3.1\pm0.6$&$5.4\pm0.6$&a\\
SVS 4-10 &$0.4\pm0.2$&$1.0\pm0.5$&$3\pm2$&$6\pm3$&b\\
\hline
\end{tabular}
\label{COiceTab}
\begin{itemize}
\item[]{References: a) \cite{Pontoppidan03a}, b) \cite{Chiar94}}
\end{itemize}
\end{flushleft}
\end{table*}

\subsection{Methanol ice}
\label{methIce}
\subsubsection{Abundance}
\begin{figure}
  \includegraphics[width=8.5cm]{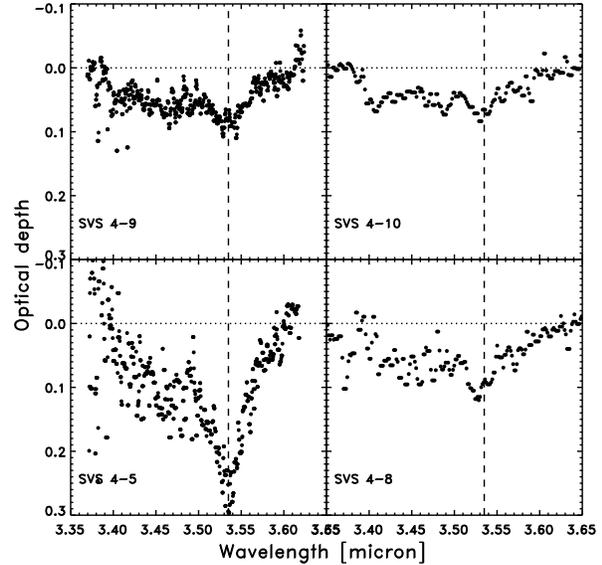}
  \caption{The four lines of sight in SVS 4 where a spectrum of good quality of the 3.53\,$\mu$m CH$_3$OH band
  could be extracted. All spectra are depicted on the same optical depth scale.
  The spectra of SVS 4-5 and 9 are new medium resolution (R=3300) observations; the other two spectra
  are taken using the low resolution mode of ISAAC yielding a resolving power of R=600. The vertical dashed line indicates 3.535\,$\mu$m. }
  \label{MethIce}
\end{figure}

\begin{figure}
  \includegraphics[width=8.5cm]{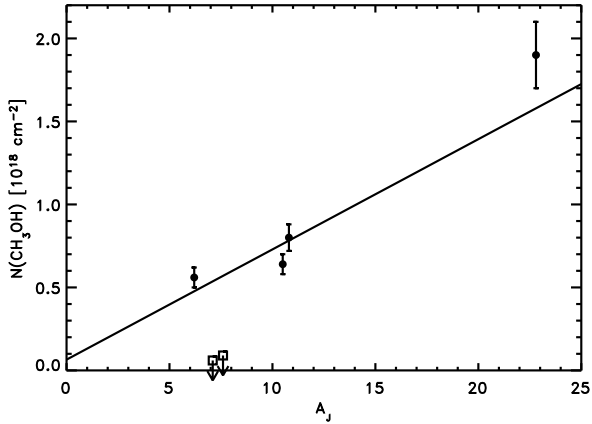}
  \caption{Relation between extinction and methanol ice column density for SVS 4. Filled circles indicate SVS 4 sources, while open squares are upper limits for the unrelated Serpens YSOs EC 90 A+B. The solid line is the best-fitting line through the SVS 4 points.}
  \label{MethAbun}
\end{figure}

One of the primary objectives of mapping ices toward SVS 4 was to explore the large excess of methanol ice known to be present 
in this region. Fig. \ref{MethIce} shows the spectra of the four lines of sight from which a high-quality 3.53\,$\mu$m methanol band could be extracted.
The remaining lines of sight do not have sufficient signal-to-noise ratios to allow a meaningful estimate of the column density of solid methanol.
In Fig. \ref{MethAbun} the relation between extinction and the column density of methanol ice is shown.
The best-fitting line to the $\tau_{\rm CH_3OH}-A_J$-relation is:
\begin{equation}
   \tau_{\rm CH_3OH} =0.06\pm 0.11+(0.07\pm 0.01)\times A_J.
\end{equation}

The relation gives a local abundance of methanol ice of $(2.5\pm 0.4) \times 10^{-5}$ relative to H$_2$. The measured threshold value is consistent with
0 mag and may therefore be identical to the water ice threshold. There is a significant scatter in the relation, which is not seen for the water ice band for the same four sources. The methanol abundances in SVS 4-5 and SVS 4-10, both of which are located in the northern part of SVS 4, are slightly higher than for SVS 4-8 and SVS 4-9. For comparison, strong upper limits on
the methanol abundance for EC 90 A and B are shown in Fig. \ref{MethAbun}. In absolute terms, the upper limits for these sources are $3-4\times 10^{-6}$. This large difference between SVS 4 and EC 90 is particularly interesting since EC 90 is located only $75\arcsec$ north of SVS 4-5. Therefore, the methanol ice must reside
in a region no more than 2\,arcminutes in extent. This region contains only SVS 4, and the outflow and envelope of SMM 4. There are thus strong indications that
the high efficiency of methanol formation in the ice mantles in SVS 4 is related to the protostellar envelope of SMM 4. 

\begin{table}
\centering
\begin{flushleft}
\caption{CH$_3$OH ice observations toward SVS 4 sources}
\begin{tabular}{lll}
\hline 
\hline 
Source & N(CH$_3$OH) & Abundance \\
              &[$10^{18}\rm cm^{-2}$]&[$10^{-5}$] w.r.t. H$_2$\\
\hline
SVS 4-5 &$1.9\pm 0.2$&$3.0\pm0.4$\\
SVS 4-8 &$0.80\pm0.08$&$2.6\pm0.4$\\
SVS 4-9 &$0.64\pm0.06$&$2.2\pm0.3$\\
SVS 4-10 &$0.56\pm0.06$&$3.2\pm0.5$\\
\hline
\end{tabular}
\label{METHiceTab}
\end{flushleft}
\end{table}

\subsubsection{Methanol band profile}
\begin{figure}
  \includegraphics[width=8cm]{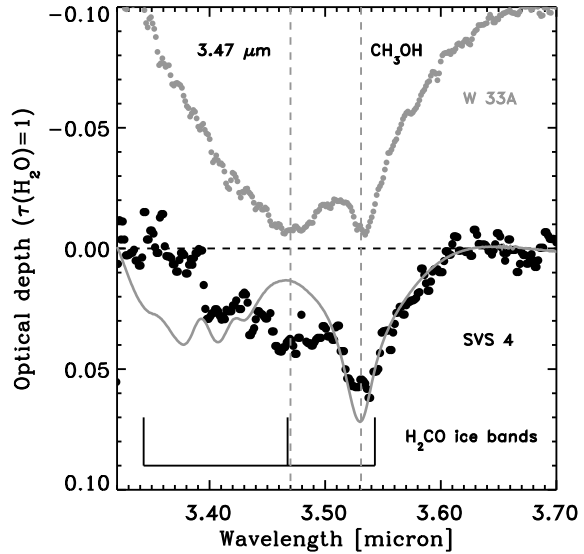} 
  \caption{Weighted average of the observed methanol bands from SVS 4-5, 4-8, 4-9 and 4-10 (black dots) compared to an ISAAC spectrum of the same region toward
  the massive YSO W 33A (grey dots). The solid line is a laboratory
  spectrum of a H$_2$O:CO$_2$:CH$_3$OH=1:1:1 mixture at 10\,K. The bands from solid H$_2$CO which may be superposed
  on the main methanol band around 3.53\,$\mu$m are indicated.}
  \label{totalMeth}
\end{figure}

The detected methanol bands toward SVS 4 sources are indistinguishable within the noise. In order to maximise the signal-to-noise, a weighted average of the four detected methanol bands has been calculated. The resulting 3.53\,$\mu$m methanol band profile is shown in Fig. \ref{totalMeth} where it is compared to an ISAAC spectrum of the well-studied methanol band toward the massive YSO W 33A \citep[e.g.][]{Dartois99}. While the methanol bands from SVS 4 are similar in shape to that of W 33A, they almost completely lack the so-called 3.47\,$\mu$m band, which is prominent toward W 33A. A laboratory spectrum of a mixture of H$_2$O:CO$_2$:CH$_3$OH=1:1:1 ices deposited at 10\,K from \cite{Ehrenfreund99} is compared to the SVS 4 methanol band. The main band at 3.53\,$\mu$m
is seen to be well matched by the laboratory profile in position and width. The secondary bands at 3.32--3.42\,$\mu$m are in general harder to match with
a laboratory spectrum. This may be due to an uncertain continuum determination in the blue wing of the band, since most of the SVS 4 sources are
very faint below 3.35\,$\mu$m. However, even the brightest source, SVS 4-9, shows no sign of excess absorption below 3.35\,$\mu$m. Some substructure absorption bands may be seen at 3.40 and 3.47\,$\mu$m, the last of which is discussed in more detail in Sec. \ref{formal}. 

\subsection{CO$_2$ ice}
\label{CO2ice}
Abundant CO$_2$ ice has been observed toward SVS 4
using ISOCAM-CVF \citep{Alexander03}. Unfortunately, many of the SVS 4 sources are blended in the CVF images, especially at the longer wavelengths, and accurate decompositions are very difficult due to the spatial undersampling of the detector.
However, our ISOCAM photometry of the high resolution broad-band imaging shows that a few sources dominate the flux at 14.3\,$\mu$m. For instance, SVS 4-5 is 6 times brighter than SVS 4-6 with which it is blended. Similarly, SVS 4-10 is 8 times brighter than SVS 4-9. In principle, the optical depth of an absorption band toward two blended sources will depend on the ratio of continuum fluxes as well as on the absolute optical depths of the two lines of sight. In particular, the depths of optically thick bands may be significantly underestimated by filling-in from another line of sight. The observed optical depth of an ice band, $\tau_{\rm obs}$, from two blended sources, $S_1$ and $S_2$, is given by:

\begin{equation}
   \exp(-\tau_{\rm obs}) = \exp(-\tau_1)[1+1/R]^{-1}+\exp(-\tau_2)[1+R]^{-1},
   \label{iceblend}
\end{equation}
where R is the ratio of the continuum flux level of $S_1$ to the continuum flux level of $S_2$. The filling-in effect is most severe if the two sources are equally bright ($R=1$) and one source has an optically thick ice band ($\tau>1$). For instance, the observed optical depth will never exceed $\ln(2)\simeq 0.7$ if the ice band is not present in one of the blended, equally bright sources.  
Therefore, if the CO$_2$ ice bands are not optically thick, it should be possible to use the ISOCAM spectra of the two bright sources to estimate reasonably accurate optical depths. 

The observed optical depths of the SVS 4-5/6 blend and the SVS 4-9/10 blend are 0.5 and 0.25, respectively. In the extreme and unlikely scenario that SVS 4-6 and SVS 4-9 have no CO$_2$ ice bands, the optical depths of SVS 4-5 and SVS 4-10 would be corrected to 0.61 and 0.29, respectively. Conversely, if the fainter sources have a very deep ice band, the actual optical depths would be corrected to
as little as 0.35 and 0.13. Since the extinction of SVS 4-5 is 5 times higher than the extinction of SVS 4-6, a large correction is expected for this line of sight due to filling-in from SVS 4-6. A smaller correction is expected for SVS 4-10, since the second source has an extinction which is only 1.7 times higher. Assuming that the local CO$_2$ abundances for the blended sources are identical, the most likely CO$_2$ ice optical depths for SVS 4-5 and SVS 4-10 are then 0.60 and 0.23, corresponding to absolute abundances of $2.0\times 10^{-5}$ and $3.5\times 10^{-5}$, respectively. Although more uncertain, the CVF spectrum of SVS 4-8 gives a  CO$_2$ ice abundance of $2\times 10^{-5}$. There are thus some indications of a slightly varying CO$_2$ ice abundance across SVS 4. It is interesting to note that the two sources with the lowest absolute abundance of CO$_2$ ice are the same sources with the highest absolute abundance of methanol ice. Although not conclusive, this may indicate a competition between the formation of 
CO$_2$ and CH$_3$OH from CO through oxidation and hydrogenation, respectively.

\subsection{Presence of formaldehyde?}
\label{formal}
\begin{figure}
  \includegraphics[width=8.5cm]{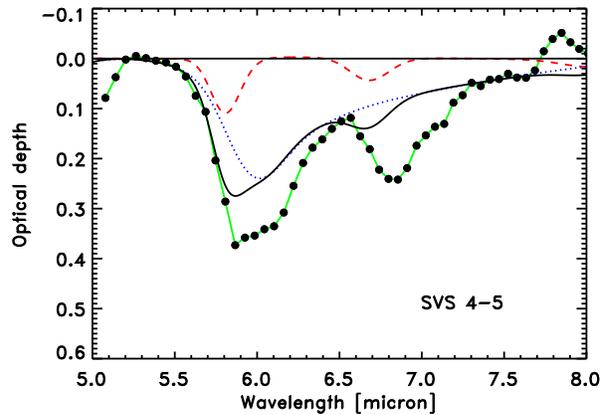}
  \caption{ISOCAM CVF spectrum of the 5-8\,$\mu$m region of the SVS 4-5/6 blend. The curves
  are laboratory spectra of pure water ice (dotted line) and pure H$_2$CO ice (dashed line). Both spectra have been convolved to
  a resolving power of $R$=35 to match the CVF spectrum. The water band has been scaled
  to match the 3.1\,$\mu$m band observed with ISAAC toward SVS 4-5. The H$_2$CO band has been scaled to the depth of the 3.47\,$\mu$m
  band. The solid line is the sum of the two laboratory spectra. }
  \label{sixpeight}
\end{figure}

The abundance of formaldehyde (H$_2$CO) ice in star forming regions has been a much debated issue. If the methanol ice has been formed through successive hydrogenation of CO, a significant abundance of formaldehyde is expected in the ice \citep{Watanabe03}. Lines of sight toward massive young stars have been found to have fairly small abundances of formaldehyde of at most 5\% relative to water \citep{Dartois99, Keane01}. 
H$_2$CO has several strong bands, but they all suffer from
blending with other strong features. The most isolated band is the $\nu_4$ mode at 3.47\,$\mu$m. There is a sharp band at this position which is common when methanol ice is abundant. However, the corresponding $\nu_1$ band of H$_2$CO at 3.54\,$\mu$m would create a significant shoulder on the red side of the 3.53\,$\mu$m methanol band, which is not clearly observed in SVS 4 (see Fig. \ref{totalMeth}). This is the same conclusion reached for other sources by \cite{Dartois99}. Formaldehyde ice has its strongest bands centered at 5.8\,$\mu$m and 6.69\,$\mu$m. These bands are
unfortunately blended with the 6.0\,$\mu$m water ice band and the 6.85\,$\mu$m band, which has still not been unambiguously identified. In Fig. \ref{sixpeight} the ISOCAM CVF spectrum of the 5-8\,$\mu$m region toward the SVS 4-5/6 blend is shown to explore whether a significant
abundance of formaldehyde is consistent with the shape of the ice bands in this region. The SVS 4-5 line of sight has the deepest ice bands as well
as the highest signal-to-noise in the CVF pointing. The ice bands are compared to laboratory spectra of pure water ice and pure formaldehyde ice 
at 10\,K. The two laboratory spectra have been convolved to the resolution of the CVF spectrum and scaled to optical depths of 3.5 and 0.1 of the 3.1\,$\mu$m water band and the 3.47\,$\mu$m formaldehyde band, respectively. This corresponds to an abundance of formaldehyde ice of $1\times 10^{-5}$ or about 10\% relative to water ice, using a band strength of $0.96\times10^{17}\rm \,cm\,molec^{-1}$ \citep{Schutte93} for the 5.8\,$\mu$m band. The CVF resolution significantly under-resolves the two formaldehyde bands, so a spectrum of higher resolution will significantly enhance the contrast of the formaldehyde to the underlying 6.0 and 6.85\,$\mu$m bands.

It is found that the water ice band cannot account for the entire 6.0\,$\mu$m band, but falls short by about 50\%. This is a well-known problem and probably indicates that one or more additional carriers contribute to the band. Typically, it is found that the 6.0\,$\mu$m band is twice as deep as it should be when comparing to the 3.1\,$\mu$m band \citep{Gibb02}. The reason that a slightly smaller discrepancy is found here may be due to veiling by SVS 4-6. According to Eq. \ref{iceblend} a blended source with a shallow ice band a factor of 10 fainter than the primary source will change the optical depth from an observed 0.37 to an intrinsic 0.42. The presence of a formaldehyde component indeed gives a better match to the blue wing of the 6.0\,$\mu$m band.
However, the 6.69\,$\mu$m band does not seem consistent with the shape of the 6.85\,$\mu$m band. We conclude that while the spectra toward the SVS 4 stars are not consistent with a high abundance of formaldehyde, an abundance of $\sim 5\%$ relative to water ice is possible. Higher resolution spectroscopy in the 5--8\,$\mu$m region is required to convincingly detect formaldehyde ice.

\begin{table}
\centering
\begin{flushleft}
\caption{Summary of ice abundances in the outer envelope of SMM 4}
\begin{tabular}{lll}
\hline
\hline
Species & Abundance & Abundance \\
&w.r.t. H$_2$O&$\times 10^{-5}$ w.r.t. H$_2$\\
\hline
H$_2$O&1&$9-17$\\
CH$_3$OH&0.28&$3$\\
Pure CO&0.3&$3$\\
CO in water&0.1-0.6&$1-5$\\
CO$_2$&0.2&$2.0-3.5$\\
H$_2$CO&$\lesssim 0.05$&$\lesssim 0.5$\\
XCN&$<0.003$&$<0.02^a$\\
\hline
\label{iceSum}
\end{tabular}
\begin{itemize}
\item[$^a$]\cite{Pontoppidan03b}
\end{itemize}
\end{flushleft}
\end{table}

\subsection{Interaction of the outflow with the ice}
SMM 4 is associated with a large bipolar outflow. Estimates of the position angle of the outflow range between 180$\degr$ \citep{Hogerheijde99} and 135$\degr$ \citep{Garay2002}. The south-eastern outflow lobe is red-shifted, while the north-western lobe is blue-shifted. An outflow is expected to evaporate at least part
of the ice mantles through shock-induced sputtering.

Observations of rotational lines of gaseous methanol have shown that the gas-phase
abundance of methanol in the SE lobe is enhanced by a factor of 40 relative to typical abundances in quiescent molecular gas of CO/CH$_3$OH$=3\times 10^{-5}$,
while the abundance in the NW lobe is enhanced by a factor of 330 \citep{Garay2002}. It is common to see such enhanced abundances in the gas phase of methanol toward bipolar outflow sources. The most favourite interpretation is that the methanol originates in icy mantles, which have been shock-evaporated by the outflow. However, it is important to note that the peak column density of gas-phase methanol toward the SE lobe of SMM 4 is only $1.4\times 10^{15}\rm\, cm^{-2}$, which translates to a gas-to-solid ratio for methanol of $\sim 2\times 10^{-3}$. Consequently, even in lines of sight with strong gas-phase enhancement, the vast majority of the methanol molecules remain as ice in the grain mantles. It is shown in Sec. \ref{methIce} that the abundance of methanol ice is enhanced by an order of magnitude in the SMM 4 outflow region. If the grain mantles in the specific case of the SMM 4 outflow were to have a typical abundance of methanol of less than a few percent relative to water ice rather than the observed 25\%, the line of sight gas-phase enhancement of methanol would be less than a factor of 4. In this case, it is doubtful that the outflow would have been recognised as having an enhanced gas-phase methanol abundance. Extrapolating this result
to other protostellar outflows with strongly enhanced gas-phase methanol abundances could indicate that the ice-phase methanol abundance
is commonly enhanced in protostellar envelopes. In most chemistry models, the ice mantles are assumed to form during the pre- and protostellar stages, either by direct freeze-out of molecules from the gas (e.g., CO) or grain-surface chemistry of accreted species (e.g., H$_2$O from accreted O).  \citep{Bergin98} proposed an alternative scheme in which ice mantles form in the wake of low-velocity shocks due to the outflow impacting on the envelope. At the high shock temperature, most of the gas-phase oxygen is driven into H$_2$O, resulting in very high H$_2$O ice abundances if all of this H$_2$O freezes out before it is chemically altered into other species in the cold post-shock gas. Such a scenario could be tested by comparing high-resolution maps of gas-phase H$_2$O with ice maps of stars behind outflow lobes such as presented here.

\subsection{Absolute ice abundances and gas-phase depletion}

The abundances relative to H$_2$ of the different ice species observed in the outer envelope of SMM 4
are summarised in Table \ref{iceSum}. The species constitute all of the most abundant products of surface oxygen chemistry according to current ice mantle models. It is therefore of interest to consider the abundances in relation to the gas phase components of the protostellar envelope and surrounding dense cloud. For instance, since the abundance of chemical products believed to originate in CO frozen out from the gas is very high, a considerable amount of the CO will be bound in the form of other molecules once the ice is returned into the gas phase upon warm-up. This will to some extent invalidate
gas phase abundances determined relative to CO even in warm gas. In other words, contrary to the gas phase, CO is a chemically active molecule in the solid phase. Toward SVS 4, the abundance of molecules formed from CO (CO$_2$, CH$_3$OH, H$_2$CO) is up to $7\times 10^{-5}$. This is up to 2 times as much as the CO ice remaining in the grain mantles. Therefore, in some cases, only 1/3 of the CO originally frozen out is returned to the gas phase as CO. The total abundance of
frozen-out CO molecules, including those which have hydrogenated or oxidised to other molecules is up to $1.4\times 10^{-4}$ relative to H$_2$, corresponding to a 
high depletion from the gas-phase.

Another interesting number to mention is the total abundance of oxygen observed in the ice. For SVS 4, 
it is as much as $2.7\times 10^{-4}$ with respect to H$_2$ and perhaps as much as $3.5\times 10^{-4}$ toward SVS 4-12. In comparison, the gas phase abundance of atomic oxygen in the diffuse interstellar medium within 500\,pc of the Sun is $\rm{O/H} = (3.19\pm0.14)\times 10^{-4}$ \citep{Meyer98}. Thus, about half of the available oxygen is bound in ice in the dense parts of the Serpens core. A similar conclusion holds for carbon. The ice abundance and composition may therefore be expected to vary considerably in regions with significantly different abundances of heavy elements. One example may be in star-forming regions in the Large Magellanic Cloud, which is known
to have a sub-solar abundance of heavy elements.

\section{Conclusions}

We have presented VLT-ISAAC $L$-band spectroscopy of most of the stars in SVS 4, a dense cluster of low-mass YSOs located near the class 0 protostar, SMM 4. These observations have been coupled with archival data from ISO and UKIRT to produce a detailed view on the distribution
of ices in the immediate environment of a very young protostar. The main conclusions are:
\begin{itemize}
\item It is shown that SVS 4 is likely located inside the south-eastern part of the outer envelope of SMM 4. The cluster members are distributed evenly throughout the envelope such that the lines of sight toward some stars probe the entire depth of the envelope. This is in particular true for SVS 4-12.
\item A $30\arcsec\times 45\arcsec$ map of the water ice abundance with a spatial resolution of $6\arcsec$ has been constructed. It shows that water ice is distributed throughout the SVS 4 cluster with an abundance of $9\pm1\times 10^{-5}$ relative to molecular hydrogen. The
abundance of water ice rises sharply to at least $1.7\times 10^{-4}$ toward SVS 4-12, which is the line of sight probing closest to the center of SMM 4. The ice along this line of sight accounts for up to 50\% of the available oxygen in the interstellar medium. This is a 
clear indication of an enhanced efficiency of ice formation in the inner parts of protostellar envelopes. 
\item Along with the water ice, methanol ice is present in SVS 4 with a fairly constant high local abundance of $2.5\pm0.4 \times 10^{-5}$ or 28\% relative to water ice. The methanol ice is confined to SVS 4, and the upper limit to the methanol ice abundance just $75\arcsec=19\,000$\,AU away from the center of SVS 4 
is $<3\times 10^{-6}$ or $<5\%$ relative to water ice. This testifies to a strongly enhanced efficiency of the formation of solid methanol in the envelope of SMM 4.
\item Other ice species show normal abundances relative to water ice. There is, however, tentative evidence for a varying abundance of CO$_2$ ice across SVS 4.
\item Mapping of ices at a spatial resolution comparable to that of gas-phase emission observations is now possible with current instrumentation on 8\,m class telescopes. Due to the fortunate alignment of a dense cluster of young stars with a protostellar envelope it was possible to achieve the high spatial resolution of $6\arcsec$. However, a spatial resolution of $30-60\arcsec$ is possible in the more general case.   
\item Maps of the abundances of H$_2$O, CH$_3$OH and CO ices relative to the column density of refractory dust or H$_2$ are potentially powerful probes of the physical history of molecular cores. Further ice mapping is essential to explore the possibilities of using ice as a physical probe. Additionally, it is important to couple maps of ices with corresponding gas-phase maps and models of the density and temperature of the dust to obtain a complete picture of the chemistry in particular of the saturated species. Indeed, about half of all molecules apart from H$_2$ not bound in refractory dust is in the form of a chemically active ice for most of the lifetime of a dense core. The Spitzer Space Telescope allows mapping at higher sensitivities in ice species not observable from the ground, such as CO$_2$ and CH$_4$.

\end{itemize}

\acknowledgements{This research is supported by a PhD grant from the Netherlands Research School for Astronomy (NOVA) and by a NWO Spinoza grant. The authors wish to thank Jens Knude for providing us with a distance estimate to Serpens and Jes J{\o}rgensen for running a physical model of SMM 4. The ISOCAM data presented in this paper were analysed using "CIA", a joint development by the ESA Astrophysics division and the ISOCAM Consortium. The ISOCAM Consortium is led by the ISOCAM PI. C. Cesarsky.}
\bibliographystyle{aa}
\bibliography{1276}
\end{document}